# Capping-induced suppression of annealing in $Ga_{1-x}Mn_xAs$ epilayers

M. B. Stone, K. C. Ku, S. J. Potashnik, B. L. Sheu, N. Samarth and P. Schiffer*

*Department of Physics and Materials Research Institute, Pennsylvania State University, University Park, PA 16802*

## ABSTRACT

We have studied the effects of capping ferromagnetic $Ga_{1-x}Mn_xAs$ epilayers with a thin layer of undoped GaAs, and we find that even a few monolayers of GaAs have a significant effect on the ferromagnetic properties. In particular, the presence of a capping layer only 10 monolayers thick completely suppresses the enhancement of the ferromagnetism associated with low temperature annealing. This result, which demonstrates that the surface of a $Ga_{1-x}Mn_xAs$ epilayer strongly affects the defect structure, has important implications for the incorporation of $Ga_{1-x}Mn_xAs$ into device heterostructures.

*schiffer@phys.psu.edu

Ferromagnetic semiconductors, in which magnetic ions incorporated into a host semiconductor lattice spontaneously align in a collectively ordered state mediated by free carriers [1,2], have become a topic of increasingly intense interest. In strong contrast to conventional metallic ferromagnets such as Fe, the low free carrier densities in semiconductors open up the possibility of "tuning" the magnetism by modulating the free carrier density either optically, electrically, or by co-doping. Such materials are hence critical to realizing semiconductor-based "spintronic" heterostructure devices [3] that would integrate magnetic properties with conventional semiconductor functionality.

Although ferromagnetism has been observed in several classes of magnetic semiconductors, the "canonical" ferromagnetic semiconductor $Ga_{1-x}Mn_xAs$ [4] is of particular importance. $Ga_{1-x}Mn_xAs$ provides a model system in which a carrier-mediated ferromagnetism persists above $T$ = 100 K [5,6,7], and recent studies consistently yield ferromagnetic transition temperatures ($T_c$) in the range 110 K < $T_c$ < 172 K [8,9,10,11]. A detailed theoretical [12,13,14,15,16,17] and experimental [7,18,19,20,21] understanding of ferromagnetism in $Ga_{1-x}Mn_xAs$ has been developing along several fronts. While it is recognized that the ferromagnetism in $Ga_{1-x}Mn_xAs$ originates from the Mn ions both acting as acceptors and providing a local moment, recent studies clearly indicate that point defects -- predominantly Mn interstitials ($Mn_I$) -- play a critical role in determining the physical behavior [19]. Such defects compensate the substitutional Mn acceptors and in the case of $Mn_I$, may reduce the ferromagnetic moment through antiferromagnetic coupling to the substitutional Mn spins [22]. Low temperature anneals have been shown to increase the carrier concentration and enhance $T_c$ [6,7,9-11,23]. This effect is particularly pronounced in thin epilayers with thicknesses of order 50 nm or less, in which $T_c$ of up to 160 K has been obtained [9-11].

We report a study of the effects of annealing on $Ga_{1-x}Mn_xAs$ epilayers that are capped by a thin layer of GaAs. Although the capping layer is typically less than ~5% of the thickness of the ferromagnetic epilayer, we find that its presence significantly suppresses $T_c$ in the as-grown samples and also reduces the physical changes induced by annealing. The effect on annealing increases with the capping layer thickness, and a 10 monolayer (ML) cap almost completely eliminates the effects of annealing. These results indicate that diffusion of defects to the free surface of $Ga_{1-x}Mn_xAs$ is an essential component of the effects of annealing, and they set important constraints on the eventual inclusion of $Ga_{1-x}Mn_xAs$ into heterostructure-based devices.

These studies focus on two series of samples (series A and series B). Samples within each series were grown with variable thickness GaAs capping layers and two different thicknesses of $Ga_{1-x}Mn_xAs$ epilayers (50 nm and 15 nm thick). Data are shown primarily from the 50 nm epilayers, but the results were qualitatively consistent with those from the 15 nm samples. All samples were grown on (001) semi-insulating, epi-ready GaAs substrates using growth conditions similar to those described elsewhere [6,]. The $Ga_{1-x}Mn_xAs$ epilayers were deposited on a buffer structure that consisted of a high temperature GaAs buffer grown at standard conditions, followed by a low temperature GaAs buffer grown at 250°C. A clear (1x2) reconstruction was observed in the reflection high electron energy diffraction (RHEED) during the growth. The GaAs capping layers with thickness ($t$) in the range $t$ = 2-15 ML were deposited by simply closing the Mn shutter after the $Ga_{1-x}Mn_xAs$ growth was completed. We studied both as-grown samples and samples which were annealed at 250 ºC for two hours in a high purity nitrogen atmosphere (99.999%) flowing at 1.5 standard cubic feet per hour. Magnetization measurements were performed in-plane using a commercial SQUID magnetometer in a magnetic field of $H = 0.002$ T after cooling in a field of $H = 1$ T [24]. Room temperature x-ray diffraction measurements were performed on the 50 nm samples using a commercial Cu-K$\alpha$ diffractometer employing a double bounce monochromator. The room temperature carrier concentration was obtained through Raman spectroscopy measurements [25] for the 50 nm series A samples. Electron microprobe analysis (EMPA) [6] indicated Mn concentrations of $x = 0.090(1)$ and $x = 0.057(3)$ in the as-grown uncapped samples from series A and B respectively.

Figure 1(a) and 1(b) depict the magnetization and resistivity as a function of temperature for as-grown and annealed 50 nm thick samples, both uncapped and capped with 10 ML of GaAs. The uncapped samples display typical behavior for both the annealed and the as-grown samples [6,7,10]. The addition of the GaAs capping layer, however, significantly reduces the $T_c$ of both the annealed and as-grown samples and also decreases the low temperature limiting magnetization by ~25 % (~15%) for annealed (as-grown) samples. We examine the development of these two effects in figures 2(a) and 2(b), which summarize the magnetic properties of the 50 nm thick epilayers [26] as a function of the thickness of the GaAs capping layer for both annealed and as-grown samples (averaging the data from growth series A and B). As seen in the figure, there is a reduction in the ferromagnetic transition temperature with increasing thickness of the capping layer for both as-grown and annealed samples. The effect is more dramatic in the annealed samples, and the difference between the annealed and unannealed samples decreases to within the scatter of the data by $t = 10$ ML. The $T = 10$ K magnetization also decreases with increasing $t$ for both annealed and as-grown samples [27]. Again, the dependence on $t$ is more dramatic in the annealed samples, and the difference between the annealed and unannealed samples decreases to within the scatter of the data for $t \geq 10$ ML.

The greatest impact of capping appears to be the suppression of the effects of annealing, and, because each series was grown using the same growth conditions, it is natural to attribute this to a suppression of the annealing-induced changes in the point defects in the ferromagnetic layer. This attribution is supported by the data in figure 2(c), in which we show the lattice constant of as-grown and annealed samples as a function of $t$. The lattice constant of the as-grown samples remains relatively constant with the addition of the GaAs capping layers. The annealed samples have markedly smaller lattice constants without a capping layer [7], but, with increasing $t$, the lattice constants of the annealed samples increase and approach those of the as-grown samples for the thickest capping layers.

The capping-induced changes in the effect of annealing are also reflected in the temperature dependence of the resistivity shown in figure 1(b) and the room temperature carrier concentration and conductivity summarized in figures 3(a) and 3(b). In uncapped samples, annealing increases the carrier concentration dramatically, and the resistivity correspondingly decreases compared to the as-grown values [7]. These differences decrease, however, with increasing $t$ as in the case of the magnetic and lattice properties. The similarities between these data and those in figure 2 suggest that the capping-induced changes in the magnetic properties are associated with the carrier concentration, since ferromagnetism in this material is strongly dependent on the density of free carriers.

We now consider how the capping layer is affecting the annealing-induced changes in defect structure. Yu *et al.* [18] have shown that one of the primary effects of annealing is to reduce the number of interstitial Mn ions. Surface analysis and modeling of the time dependence of annealing suggest that interstitials migrate to the free surface of the epilayer during the annealing process [28], providing a natural explanation for the recently observed dependence of $T_c$ on the $Ga_{1-x}Mn_xAs$ epilayer thickness [10, 29]. This scenario also provides a plausible model within which to understand the effects of capping. Upon annealing a capped sample, the diffusion of $Mn_I$ into the capping layer should n-dope the capping layer (in which there are no substitutional Mn ions providing holes). The resultant p-n junction created at the interface between the capping layer and the hole-doped $Ga_{1-x}Mn_xAs$ layer would then provide a repulsive Coulomb barrier limiting the number of $Mn_I$ which could diffuse to the surface from the $Ga_{1-x}Mn_xAs$ and thus limit the effects of annealing associated with removal of the $Mn_I$ [28]. Such a process would also explain why the as-grown capped samples have a slightly reduced $T_c$ and low temperature moment, since there presumably is some diffusion of $Mn_I$ to the free surface while a sample cools immediately after completion of growth.

The suggestion that the effects of capping are associated with $Mn_I$ being unable to migrate to the surface is further supported by EMPA measurements on un-capped samples. These measurements indicate an apparent *rise* in the Mn concentration of uncapped samples upon annealing from $x = 0.090(2)$ to $x = 0.123(1)$ for growth A and from $x = 0.057(3)$ to $x = 0.074(3)$ for growth B. Because of the finite penetration depth of the probing electrons, the EMPA measurements are most sensitive to the material near the sample surface, and this result is thus consistent with $Mn_I$ diffusing toward the surface of the sample.

Our results provide strong, albeit indirect, evidence that diffusion of defects to the free surface is an important source of the benefits of annealing $Ga_{1-x}Mn_xAs$, supporting similar conclusions based on other data [28]. Perhaps more importantly, the results imply that $Ga_{1-x}Mn_xAs$ epilayers incorporated in heterostructures will always behave differently from single epilayer samples, and indeed there are examples of exactly this effect in the recent literature [11,30]. While it has been previously recognized that defects are a significant component of the physics of $Ga_{1-x}Mn_xAs$, the capping data imply that the controlling defect structure is affected not only by growth and annealing conditions but also by adjacent semiconductor layers. Since this material is of interest for spintronic devices composed of epitaxially grown heterostructures, these data have potentially important implications for the design of such devices.

The authors would like to thank T. Dietl, A. MacDonald, and B. Gallagher, for valuable discussions. We also thank M. J. Seong and A. Mascarenhas for Raman measurements. N.S. and K.C.K. were supported by ONR N00014-99-1-0071 and –0716, and DARPA/ONR N00014-99-1-1093. M.B.S., S.J.P., B.L.S. and P.S. were supported by DARPA N00014-00-1-0951 and NSF DMR-0101318.

**Figure 1.** Magnetization (a) and resistivity (b) as a function of temperature for annealed and as-grown 50 nm thick $Ga_{1-x}Mn_xAs$ samples. Results shown are from series A samples with (closed symbols) and without (open symbols) 10 ML thick GaAs capping layers. The addition of the capping layer is shown to reduce $T_c$ for both as-grown and annealed samples. In addition, the capping layer increases the resistivity for both as-grown and annealed samples over a significant portion of the temperature range probed.

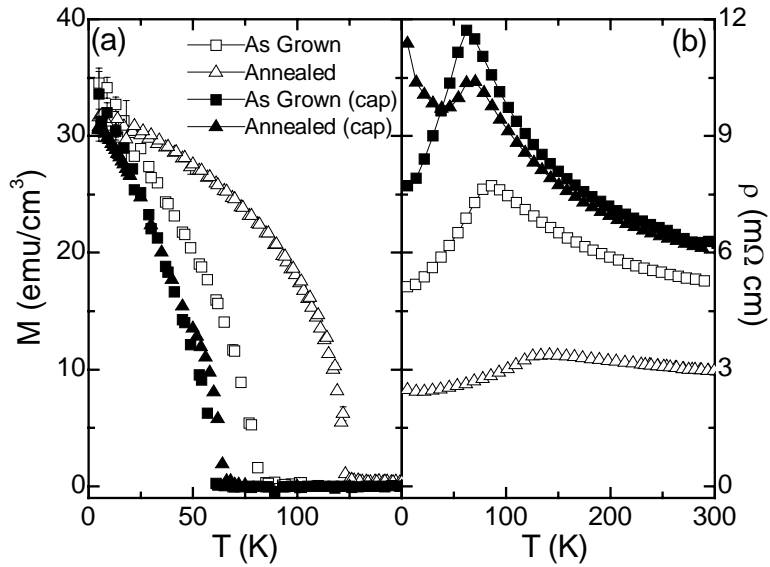

**Figure 2** Summary of the (a) Curie temperature, $T_c$, (b) magnetization at T=10 K, and (c) the room temperature relaxed lattice constant as a function of the number of capping monolayers of GaAs for 50 nm thick samples of $Ga_{1-x}Mn_xAs$. Data are shown for both as-grown and annealed samples with results averaged from measurements performed on both series A and series B sample growths.

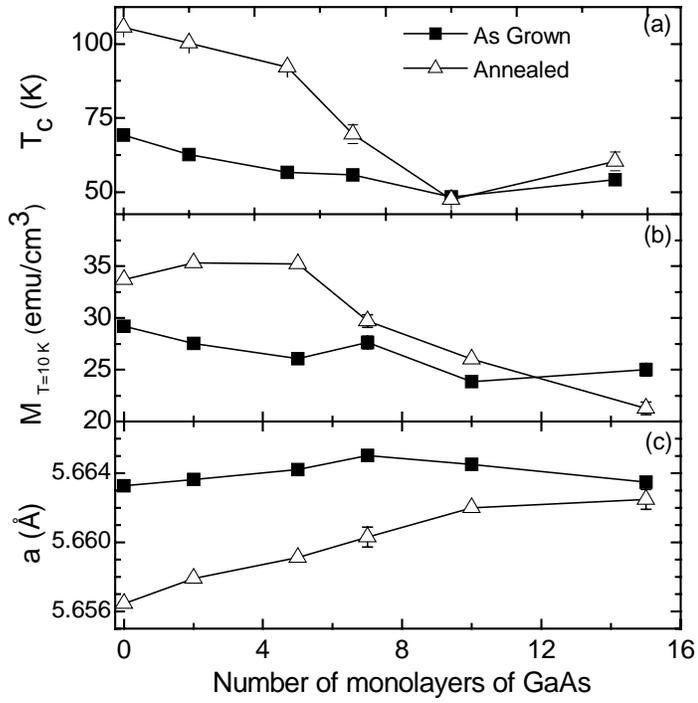

**Figure 3** (a) Carrier concentration, $p$, and (b) conductivity, $\sigma$, at $T = 300$ K as a function of the number of GaAs capping monolayers for 50 nm thick $Ga_{1-x}Mn_xAs$ epilayers from growth series A, both as-grown and annealed samples.

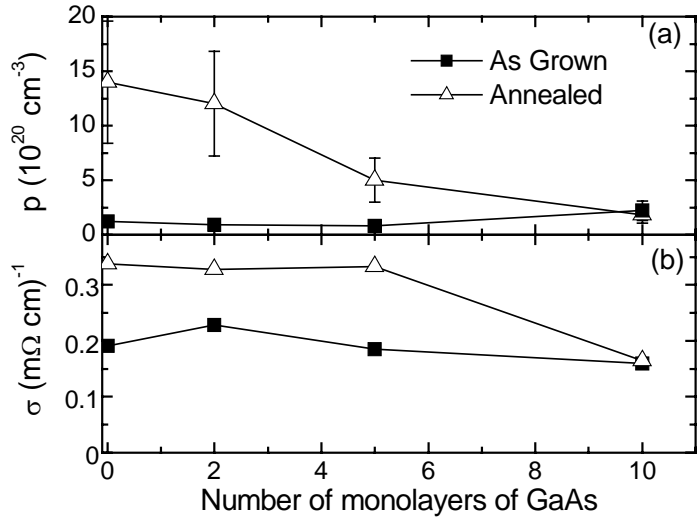